\PassOptionsToPackage{unicode}{hyperref}
\PassOptionsToPackage{hyphens}{url}
\documentclass[letterpaper]{article}

\usepackage{xcolor}
\usepackage{amsmath,amssymb}
\usepackage{iftex}
\ifPDFTeX
  \usepackage[T1]{fontenc}
  \usepackage[utf8]{inputenc}
  \usepackage{textcomp} 
\else 
  \usepackage{unicode-math} 
  \defaultfontfeatures{Scale=MatchLowercase}
  \defaultfontfeatures[\rmfamily]{Ligatures=TeX,Scale=1}
\fi
\usepackage{lmodern}
\ifPDFTeX\else
\fi
\IfFileExists{upquote.sty}{\usepackage{upquote}}{}
\IfFileExists{microtype.sty}{
  \usepackage[]{microtype}
  \UseMicrotypeSet[protrusion]{basicmath} 
}{}
\makeatletter
\@ifundefined{KOMAClassName}{
  \IfFileExists{parskip.sty}{%
    \usepackage{parskip}
  }{
    \setlength{\parindent}{0pt}
    \setlength{\parskip}{6pt plus 2pt minus 1pt}}
}{
  \KOMAoptions{parskip=half}}
\makeatother
\usepackage{longtable,booktabs,array}
\usepackage{calc} 
\usepackage{etoolbox}
\makeatletter
\patchcmd\longtable{\par}{\if@noskipsec\mbox{}\fi\par}{}{}
\makeatother
\IfFileExists{footnotehyper.sty}{\usepackage{footnotehyper}}{\usepackage{footnote}}
\makesavenoteenv{longtable}
\usepackage{bookmark}
\usepackage[margin=1.5in]{geometry}
\usepackage[hang,flushmargin]{footmisc}
\usepackage{titling}
\IfFileExists{xurl.sty}{\usepackage{xurl}}{} 
\makeatletter
\@ifundefined{xmpquote}{}{}
\makeatother
\hypersetup{
  hidelinks,
  pdfcreator={LaTeX via pandoc}}

\title{Pricing the Unpriced Asset: A Standards-Based Method for Valuing Enterprise Data under IAS 38 and IAS 2
 \\[1em]
       \large Introducing D-Val and A-Val as a Cost-Floor Model for Data Asset Valuation across Intangibles and Inventory}

\author{Natasha E. Blycha,\quad James Myint,\quad Dean Arden,\quad Michael Small,\\ Conor Blycha, \quad Schellie-Jayne Price,\quad Steve Bailey,\\ Ryan Feng,\quad Rachael Johnson,\quad Adam Myers}

\date{June 2026} 

\begin{document}
\maketitle
\textbf{Abstract}

The recognition and measurement of data assets under current accounting
standards presents significant theoretical and practical challenges.
While International Accounting Standard 38 (IAS 38) provides a framework
for intangible asset recognition, data assets frequently fail to meet
capitalisation criteria due to difficulties in demonstrating
separability, establishing reliable cost measurement, and proving
probable future economic benefits.

The widespread failure to easily and reliably value data causes mispricing and allocative distortions across data and artificial intelligence markets.

This paper introduces a two-layer valuation progression for
authenticated data assets, that is, datasets that have met IAS 38
recognition criteria through established legal provenance and
contractual boundaries. The first layer, D-Val, is the auditable
cost-basis valuation consistent with IAS 38 as currently applied. D-Val
is defined as \(\text{D-Val}=Cp \times Av^t\), where $Cp$ is the reliably measurable
production cost and $Av^t$ is the appreciation or depreciation factor
applied over time. Under prevailing interpretations of IAS 38, $Av$ is
constrained to values less than or equal to one absent an active market
revaluation, rendering D-Val a strictly cost-less-amortisation figure.
The second layer, A-Val, is a theoretically grounded commercial
valuation that incorporates scarcity, rivalry, completeness, accuracy,
and explicit premia for provenance authentication and independent audit.
A-Val is not auditable as fair value under current practice but serves
as a defensible commercial valuation during the period before active
markets for authenticated data assets mature.

The progression is therefore \(\text{D-Val} \leq \text{A-Val} \leq \text{Market value}\), with A-Val
bounded below by D-Val through a cost floor provision. Three detailed
worked examples demonstrate the methodology across retail, mining, and
healthcare sectors, with A-Val central estimates ranging from 1.00x
(floored at D-Val) to 2.97x of D-Val. The methodology incorporates
provisional premia for authentication (20\%) and audit verification
(20\%) derived from related certification markets and economic theory.
Parameter assumptions requiring empirical validation as authenticated
data markets mature are explicitly identified, providing a foundation
for iterative refinement as transaction data becomes available.

\newpage
\section*{Table of Contents}\label{toc}

\hyperref[introduction]{1. Introduction}

\hyperref[theoretical-framework]{2. Theoretical Framework}

\hyperref[data-versus-authenticated-data-assets]{\hspace{1em}2.1 Data Versus Authenticated Data Assets}

\hyperref[economic-characteristics-of-data-assets]{\hspace{1em}2.2 Economic Characteristics of Data Assets}

\hyperref[rivalry-and-excludability]{\hspace{2em}2.2.1 Rivalry and Excludability}

\hyperref[current-accounting-treatment-and-limitations]{3. Current Accounting Treatment and Limitations}

\hyperref[ias-38-recognition-criteria]{\hspace{1em}3.1 IAS 38 Recognition Criteria}

\hyperref[traditional-valuation-method-limitations]{\hspace{1em}3.2 Traditional Valuation Method Limitations}

\hyperref[multi-period-excess-earnings-method-mpeem]{\hspace{2em}3.2.1 Multi-Period Excess Earnings Method (MPEEM)}

\hyperref[market-based-approach]{\hspace{2em}3.2.2 Market-based Approach}

\hyperref[cost-approach]{\hspace{2em}3.2.3 Cost Approach}

\hyperref[the-two-layer-valuation-progression]{4. The Two-Layer Valuation Progression}

\hyperref[d-val-the-auditable-cost-basis-valuation]{\hspace{1em}4.1 D-Val: The Auditable Cost-Basis Valuation}

\hyperref[the-progression-from-d-val-to-market-value]{\hspace{1em}4.2 The Progression from D-Val to Market Value}

\hyperref[a-val-the-commercial-valuation]{\hspace{1em}4.3 A-Val: The Commercial Valuation}

\hyperref[parameter-definitions]{\hspace{2em}4.3.1 Parameter Definitions}

\hyperref[limitation-parameter-backlogging]{\hspace{2em}4.3.2 Limitation: Parameter Backlogging}

\hyperref[the-provenance-premium]{\hspace{1em}4.4 The Provenance Premium: Theoretical Foundations}

\hyperref[risk-reduction-through-information-asymmetry-mitigation]{\hspace{2em}4.4.1 Risk Reduction Through Information Asymmetry Mitigation}

\hyperref[transaction-cost-reduction-through-property-rights-clarity]{\hspace{2em}4.4.2 Transaction Cost Reduction Through Property Rights Clarity}

\hyperref[network-effects-and-market-liquidity-enhancement]{\hspace{2em}4.4.3 Network Effects and Market Liquidity Enhancement}

\hyperref[parameter-specification-and-justification]{\hspace{2em}4.4.4 Parameter Specification and Justification}

\hyperref[the-audit-premium-theoretical-foundations]{\hspace{1em}4.5 The Audit Premium: Theoretical Foundations}

\hyperref[certification-value-and-signalling-theory]{\hspace{2em}4.5.1 Certification Value and Signalling Theory}

\hyperref[agency-cost-reduction]{\hspace{2em}4.5.2 Agency Cost Reduction}

\hyperref[insurance-value-and-risk-transfer]{\hspace{2em}4.5.3 Insurance Value and Risk Transfer}

\hyperref[empirical-evidence-from-related-markets]{\hspace{2em}4.5.4 Empirical Evidence from Related Markets}

\hyperref[parameter-specification-and-justification-1]{\hspace{2em}4.5.5 Parameter Specification and Justification}

\hyperref[combined-authentication-and-audit-effects]{\hspace{2em}4.5.6 Combined Authentication and Audit Effects}

\hyperref[parameter-uncertainty-and-sensitivity]{\hspace{1em}4.6 Parameter Uncertainty and Sensitivity}

\hyperref[worked-examples]{5. Worked Examples}

\hyperref[example-1-retail-customer-transaction-database]{\hspace{1em}5.1 Example 1: Retail Customer Transaction Database}

\hyperref[example-2-mining-geological-survey-dataset]{\hspace{1em}5.2 Example 2: Mining Geological Survey Dataset}

\hyperref[example-3-de-identified-clinical-outcomes-database]{\hspace{1em}5.3 Example 3: De-identified Clinical Outcomes Database}

\hyperref[cross-example-analysis]{\hspace{1em}5.4 Cross-Example Analysis}

\hyperref[conclusion-and-implementation-guidance]{6. Conclusion and Implementation Guidance}

\hyperref[limitations-and-future-research]{\hspace{1em}6.1 Limitations and Future Research}

\hyperref[references]{References}

\newpage
\section{1. Introduction}\label{introduction}

The global digital economy has reached a critical juncture where data is
a fundamental source of competitive advantage and economic value across
virtually all industry sectors. Organisations invest billions annually
in data infrastructure, collection, curation, and analysis, yet current
accounting practices create a systematic disconnect between this
economic reality and financial reporting. Under prevailing
interpretations of International Accounting Standard 38 (IAS 38) and
equivalent standards in other jurisdictions, data development
expenditures are typically expensed as incurred rather than capitalised
as assets\footnote{Multi-client seismic data acquired by seismic
  companies and sold to multiple clients on a licensed basis is an
  example of data being recognised as a capitalised asset on the balance
  sheet.}, resulting in balance sheets that fail to reflect substantial
organisational resources.\footnote{Seventy per cent of respondents
  queried by the CFA Institute viewed intangibles such as data as the
  most valuable assets for a company but observed that existing
  accounting models do not recognise them as assets International
  Accounting Standards Board (IASB). (2022). \emph{IAS 38 Intangible
  Assets: Basis for Conclusions and Dissenting Opinions}. IFRS
  Foundation.}

The consequences of this disconnect do not stay inside the firm. Data has become the principal input to artificial intelligence (AI), and capital is flowing toward AI at a scale that makes what that input is worth a matter of broad economic and policy significance rather than internal bookkeeping. The widespread failure to value data reliably propagates as mispricing and allocative distortion across data and AI markets, steering capital toward firms and projects on the basis of figures that omit their most important resource. The phenomenon is now recognised as data market failure, in which the difficulty of accurately assessing data value makes pricing a central problem. While this paper is concerned primarily with how best to value data assets, any progress on data valuation should also assist the structural allocation of resources in the AI economy.\footnote{Deng, Xiannian and Yong Shi, ‘A Study on the Causes and Regulation of Data Market Failure’ (2025) 266 Procedia computer science 1220; Acemoglu, Daron et al, ‘Too Much Data: Prices and Inefficiencies in Data Markets’ (2022) 14(4) American economic journal. Microeconomics 218} 

The conservative accounting treatment by IAS 38 and its application
creates further problems in practice. First, it understates organisational value by
excluding significant economic assets from financial statements. Second,
it distorts investment decisions by failing to provide management with
accurate information about data asset performance and returns. Third, it
creates artificial distinctions between economically identical data
assets based solely on their acquisition method: data acquired through
business combinations may be recognised at fair value under IFRS 3,
while internally developed data meeting identical economic criteria is
expensed. Fourth, it potentially creates inequality in information
markets, by allowing organisations with proprietary models and
proprietary access to data, to value intangibles in a more robust manner
to broader markets, fundamentally causing distortions between private
and public markets.

Recent technological developments enable data to be authenticated
through legal and technical mechanisms that establish clear ownership
rights, usage permissions, and chain of custody. Such authenticated data
assets appear to meet the separability and contractual-legal criteria
required for recognition under IAS 38. Standard practice has not yet
adopted capitalisation, due in part to a lack of familiarity with this
emerging asset class and to the remaining challenge of establishing
reliable fair value measurements (Rajuroy, 2021).

Real world technological developments reflect real world market demand.
AI and data companies and their executives, including Oracle's Executive
Chairman and CTO, Larry Ellison, have acknowledged the technical
requirement for data as a feedstock for AI model training and the
coinciding commercial opportunity in monetising proprietary enterprise
data.\footnote{\url{https://www.ibtimes.co.uk/larry-ellison-says-ai-race-will-led-those-access-private-enterprise-data-1774101}}
Anecdotal evidence indicates that data licence fees in the eight figures
per year are being exchanged for single datasets.

This paper addresses the gap between traditional accounting standards
and emerging market demand by articulating a two-layer valuation
progression for authenticated data assets. The first layer, D-Val, is
the auditable cost-basis valuation consistent with IAS 38 as currently
applied. The second layer, A-Val, is a commercial valuation that
incorporates data-specific characteristics including scarcity,
completeness, accuracy, and legal provenance. The progression between
the two layers maps onto a clear commercial narrative: authentication
creates a separable asset by isolating the dataset from the broader
operational and informational substrate and wrapping it in contractual
boundaries; this separation in turn creates the preconditions for an
active market to form; once transactions begin to occur at scale,
independent audit and verification enable revaluation; and at market
maturity, A-Val becomes the candidate methodology for auditable fair
value measurement. The present paper sits between the first and second
steps of that progression, providing a methodology that is commercially
useful today while awaiting market maturation.

The methodology goes some way towards addressing the causality dilemma
of data markets. Data must be valued to establish the utility of a
marketplace, yet without an active marketplace it is difficult to
determine value, because the value of data is highly context and
use-case specific. By providing a conservative, auditable floor (D-Val)
and a defensible commercial estimate (A-Val), the progression creates a
pricing vocabulary that can seed market formation without overreaching
the current evidentiary base.

The paper proceeds as follows. Section 2 examines the theoretical
foundations for data asset recognition, distinguishing raw data from
authenticated data assets. Section 3 reviews current accounting
treatment and identifies specific limitations preventing capitalisation.
Section 4 presents the two-layer progression, defining D-Val and A-Val,
providing theoretical justification for both provenance and audit premia
grounded in property rights theory (Coase, 1960; Barzel, 1982),
signalling theory (Spence, 1973), and agency cost frameworks (Jensen \&
Meckling, 1976). Section 5 provides three detailed worked examples
demonstrating practical application and identifies limitations in the
approach. Section 6 concludes with implementation guidance and future
research directions.

\section{2. Theoretical Framework}\label{theoretical-framework}

\subsection{2.1 Data Versus Authenticated Data
Assets}\label{data-versus-authenticated-data-assets}

A fundamental distinction exists between raw data and authenticated data
assets. Raw data comprises unorganised information bits with limited
property characteristics. Coyle and Manley (2020) characterise raw data
along economic and informational dimensions, identifying non-rivalry,
externalities, and quality variability as key features. From a legal
perspective, raw data lacks the identifiability and alienability
required for property status in most jurisdictions.

In contrast, authenticated data assets represent organised datasets with
established legal boundaries, documented ownership rights, and traceable
provenance. Authentication transforms data through several mechanisms:

\begin{itemize}
\item
  \textbf{Contractual definition.} Legal instruments define precise
  dataset boundaries, distinguishing specific data from general
  information flows.
\item
  \textbf{Rights specification.} Documented ownership, usage
  permissions, and transferability establish clear property rights.
\item
  \textbf{Provenance tracking.} Chain of custody documentation enables
  verification of data origins and subsequent handling.
\item
  \textbf{Quality certification.} Independent verification of
  completeness, accuracy, and consistency.
\end{itemize}

These authentication characteristics align authenticated data assets
with IAS 38 recognition criteria. The standard requires intangible
assets to be identifiable, controlled by the entity, capable of reliable
cost measurement, and expected to generate future economic benefits.
Authentication establishes identifiability through separability, in that
the dataset can be sold, licensed (exclusively or non-exclusively), or
exchanged independently, and through contractual-legal rights that
distinguish the asset from goodwill. Control is determined by
authentication, specifically contractual definition via a suitably
established software platform, rights specification, and provenance
tracking, which together delineate rights in the data asset as against
other stakeholders. Authentication and contractual wrapping in this
manner finally allows the data to be licensed, sold, or exchanged with
other persons for value, and thus to derive economic benefits.

A scoping caveat is warranted. Where data is held for sale in the
ordinary course of business, it should be accounted for, but under the
trading stock (inventory) rather than as an intangible asset. The
distinction turns on the entity\textquotesingle s intention and the
nature of the holding. Data held for use, licensing, or long-term
exploitation falls within the scope of this paper and is analysed under
IAS 38. Data held for sale as trading stock, whilst still potentially
under accounted for, is subject to different accounting and tax
treatment and is outside the scope of this analysis.

\subsection{2.2 Economic Characteristics of Data
Assets}\label{economic-characteristics-of-data-assets}

Data assets exhibit distinctive economic properties that influence
valuation methodology. Understanding these characteristics provides
theoretical foundation for the A-Val formula parameters discussed in
Section 4.

\subsubsection{2.2.1 Rivalry and
Excludability}\label{rivalry-and-excludability}

Raw data is inherently non-rivalrous.\footnote{Although in the case of
  licensing, an exclusive data licence is rivalrous and in certain cases
  the value of a non-exclusive data licence may erode as each further
  non-exclusive licences is granted.} One party's use does not prevent
simultaneous use by others. However, authenticated data assets can be
rendered rivalrous through contractual restrictions and technological
access controls. The degree of rivalry varies by data type and
application context. Healthcare data may exhibit near-perfect rivalry
due to privacy regulations and competitive sensitivity, while public
meteorological data demonstrates minimal rivalry despite substantial
economic value.

This variable rivalry creates valuation challenges absent in traditional
assets. The A-Val methodology addresses this through a rivalry factor
($\beta$) ranging from 0 (non-rivalrous) to 1 (perfectly rivalrous), with
systematic determination based on data type, sensitivity, and regulatory
context. Rivalry only directly affects the commercial layer (A-Val) and
does not enter the auditable cost-basis layer (D-Val).

\section{3. Current Accounting Treatment and
Limitations}\label{current-accounting-treatment-and-limitations}

\subsection{3.1 IAS 38 Recognition
Criteria}\label{ias-38-recognition-criteria}

International Accounting Standard 38 establishes three fundamental
criteria for intangible asset recognition. First, the asset must be
identifiable, either through separability or through arising from
contractual or legal rights. Second, the entity must control the asset
such that it can obtain future economic benefits and restrict others'
access to those benefits. Third, and as a related point to the previous
criterion, the cost of the asset must be reliably measurable.

For internally generated intangible assets, IAS 38 imposes additional
requirements. Development expenditure may be capitalised only when
technical feasibility, intention to complete, ability to use or sell,
probable future economic benefits, availability of resources, and
reliable cost measurement can all be demonstrated. Research expenditure
must be expensed as incurred.

IAS 38 further specifies measurement after recognition. The cost model
is the default: an intangible asset is carried at cost less any
accumulated amortisation and any accumulated impairment losses. The
revaluation model is only available where fair value can be determined
by reference to an active market. IAS 38 requires three conditions to be
demonstrable for a market to qualify as active:\footnote{See also IFRS
  13 for the definition of an active market and fair value measurements
  of assets; also see for conditions relating to an active market and
  determining a price with respect to the market.} the items traded must
be homogeneous, willing buyers and sellers must ordinarily be available
at any time, and prices must be available to the public. Active markets
for intangible assets are in practice rare, as most intangibles are
insufficiently homogeneous and transactions are insufficiently public to
satisfy these conditions. For authenticated data assets today, no active
market exists in the sense contemplated by IAS 38, with the consequence
that even where the recognition criteria are met, the asset must be
carried at cost less accumulated amortisation and impairment. This is
the mechanical origin of the D-Val expression introduced in Section 4.

IAS 38 was originally issued nearly three decades ago, and these
standards were developed with no conception of the technology available
today which over the past decades has fundamentally changed how value
can be reliably measured and where this value can be found. IAS 38 is a
conservative standard but simply reflecting its time.. Accounting
professionals have applied the conservative standards to modern data
assets, which has resulted in potential questions around distinguishing
data development from business development, in proving probable future
benefits without established markets, and in reliably measuring fair
value resulting in data being left off balance sheets.

\subsection{3.2 Traditional Valuation Method
Limitations}\label{traditional-valuation-method-limitations}

Existing intangible asset valuation methodologies encounter specific
challenges when applied to data assets.

\subsubsection{3.2.1 Multi-Period Excess Earnings Method
(MPEEM)}\label{multi-period-excess-earnings-method-mpeem}

The MPEEM requires isolating cash flows attributable solely to the data
asset being valued. For data assets this presents insurmountable
difficulties. Data creates value through integration with other business
assets including analytical software, human capital, customer
relationships, operational systems, and increasingly IoT devices. These
assets exhibit circular dependencies: data enhances software
effectiveness while software makes data more valuable. Attempting to
decompose this integrated value creation into independent cash flow
streams proves both practically impossible and theoretically unsound.
Authentication extracts the data from these entanglements, making it
severable and therefore amenable in principle to independent valuation.

\subsubsection{3.2.2 Market-based Approach}\label{market-based-approach}

Market-based valuation requires comparable transactions for similar
assets. While data licensing and sales occur with increasing frequency,
most transactions remain opaque, conducted through over-the-counter
arrangements with undisclosed terms. Even when transaction prices become
public, determining comparability proves challenging because the exact
terms of the contractual arrangements are often unknown (exclusive
licence, non-exclusive licence, term of licence, use or other
restrictions such as prohibitions on sublicensing) and because datasets
are heterogeneous in size, quality, fidelity, structure, completeness,
accuracy, provenance, authenticity, time duration (such as time series
data), recency (freshness), and permitted uses.

\subsubsection{3.2.3 Cost Approach}\label{cost-approach}

The cost approach is well established for other internally generated
intangible assets. Capitalised software development costs, patent
registration costs, and development-phase expenditure under IAS
38\textquotesingle s criteria are all carried at cost less accumulated
amortisation and impairment. Entities do not set up their accounting
systems to capture this information, presumably because the current
standard selects against it. The reliable-cost-measurement test in these
contexts is satisfied through project-based cost attribution in the
entity\textquotesingle s accounting systems, supported by timesheet
allocation, infrastructure cost allocation, and documented
capitalisation policies. Authenticated data assets are amenable to the
same cost-tracking methodology: where an entity operates project-coded
accounting for data asset development, direct costs (personnel, storage,
tooling, data acquisition) and capitalisable overheads can be attributed
to the specific dataset with the same degree of reliability that is
already accepted for software capitalisation.

The cost approach under IAS 38 values an intangible asset at the cost of
its development or acquisition, less accumulated amortisation and
impairment. For many authenticated data assets this produces a
demonstrably conservative figure, since it does not capture scarcity,
rivalry, quality, or the premium that authentication itself is likely to
command in a mature market. Further, the standard requires expensing
research costs, creating practical difficulties in determining which
costs qualify for capitalisation when data collection and organisation
occur continuously rather than in discrete development phases.

The cost approach is nonetheless the only approach currently supported
by the recognition and measurement framework of IAS 38 in the absence of
an active market. It is this approach that D-Val formalises, and it is
the gap between this approach and economic reality that A-Val is
designed to bridge.

\section{4. The Two-Layer Valuation
Progression}\label{the-two-layer-valuation-progression}

\subsection{4.1 D-Val: The Auditable Cost-Basis
Valuation}\label{d-val-the-auditable-cost-basis-valuation}

D-Val is defined as the auditable cost-basis valuation of an
authenticated data asset, expressed as:
\[\text{D-Val} = Cp \times Av^t\]

where Cp is the reliably measurable production cost of the dataset, Av
is the appreciation or depreciation factor, and t is the elapsed time
since acquisition or last measurement. Under prevailing interpretations
of IAS 38, Av is constrained to values less than or equal to 1 in the
absence of an active market revaluation. Where the asset is being
amortised on a systematic basis, Av\^{}t represents one minus the
cumulative amortisation percentage, and any impairment losses are
applied as a further reduction.

Expressed in narrative form, D-Val is the carrying value of the asset
under IAS 38: cost of production less accumulated amortisation and
impairment. It is the number that can be placed on a balance sheet
today, subject to the entity meeting the identifiability, control,
reliable cost measurement, and probable future economic benefits tests
discussed in Section 3.1.

Three features of D-Val warrant emphasis. First, D-Val is auditable now.
The inputs (production cost, amortisation schedule, impairment
assessment) are conventional accounting determinations, supported by
existing audit methodology, and measurement of Cp relies on the same
project-based cost-tracking infrastructure that is already used for
software capitalisation under IAS 38. Second, D-Val is conservative by
design. It incorporates none of the scarcity, rivalry, quality, or
authentication premia that drive commercial value in data markets.
Third, D-Val is the appropriate floor for any reported valuation of an
authenticated data asset. As discussed in Section 4.3, A-Val is defined
to include D-Val as a structural floor, such that the commercial
valuation cannot produce a reported figure below the auditable carrying
value. D-Val therefore functions not merely as a conservative
alternative to A-Val, but as a binding lower bound embedded within it.

The definition \( \text{D-Val} = Cp \times Av^t\) is expressed in a general form to
accommodate future regulatory evolution. Once an active market for
authenticated data assets exists in the sense contemplated by IAS 38,
that is, where homogeneous authenticated datasets can be shown to trade
between willing buyers and sellers at publicly available prices, the
revaluation model would permit Av to take values greater than 1 where
the market supports it. The formula in that state of the world remains
unchanged; only the constraint on Av relaxes. The D-Val framework is
therefore stable across the progression from the current regulatory
state to a future market-supported revaluation state.

A note on scope. Although this paper concerns authenticated data assets,
the D-Val construct extends by analogy to any intangible asset
authenticated through the same mechanisms of contractual definition,
rights specification, and provenance tracking, including authenticated
Artificial Intelligence (AI) Agents constructed under equivalent legal
and contractual boundaries; the A-Val construct, by contrast, is
calibrated to dataset-specific parameters and does not extend to agents
without substantial reformulation.

\subsection{4.2 The Progression from D-Val to Market
Value}\label{the-progression-from-d-val-to-market-value}

The two quantities introduced in this paper, together with the eventual
market value, stand in the following relationship:

\[\text{D-Val} \leq \text{A-Val} \leq \text{Market value}\]

The commercial logic behind this progression is straightforward.
Authentication creates the asset by separating the dataset from the
broader operational substrate and wrapping it in contractual boundaries.
Separation is the precondition for an active market: absent separation,
there is nothing discrete to trade. Once authenticated assets begin to
be exchanged at observable prices, a corpus of transaction data
accumulates, enabling independent audit and verification of value
claims. As the transaction corpus grows, the parameters of any candidate
valuation methodology (including A-Val) can be empirically calibrated
against observed prices. At market maturity, revaluation of
authenticated data assets by reference to the active market becomes
possible under the IAS 38 revaluation model, and the commercial
valuation produced by A-Val becomes a candidate for auditable fair value
measurement.

The two-layer progression set out in this paper addresses the interval
before market maturity. During this interval, D-Val provides the
auditable number that sits on the balance sheet, while A-Val provides a
defensible commercial estimate that supports licensing negotiations,
strategic planning, and internal capital allocation decisions. The
progression is summarised in Figure 1.

{\def\LTcaptype{none} 
\begin{longtable}[]{@{}
  >{\raggedright\arraybackslash}p{(\linewidth - 6\tabcolsep) * \real{0.1816}}
  >{\raggedright\arraybackslash}p{(\linewidth - 6\tabcolsep) * \real{0.3312}}
  >{\raggedright\arraybackslash}p{(\linewidth - 6\tabcolsep) * \real{0.3312}}
  >{\raggedright\arraybackslash}p{(\linewidth - 6\tabcolsep) * \real{0.1560}}@{}}
\toprule\noalign{}
\endhead
\bottomrule\noalign{}
\endlastfoot
\textbf{Layer} & \textbf{Formula} & \textbf{Purpose} & \textbf{Auditable
today?} \\
D-Val & $Cp \times Av^t, (Av \leq 1)$& Balance sheet carrying value under IAS
38 & Yes \\
A-Val & Quality-adjusted commercial formula (Section 4.3) & Commercial
valuation for licensing, planning, and negotiation & Not yet; candidate
for auditable fair value at market maturity \\
Market value & Observed transaction price in an active market & Fair
value under IAS 38 revaluation model once active market exists & Once
market matures \\
\end{longtable}
}

\emph{Figure 1. The valuation progression for authenticated data
assets.}

A practical constraint links the two layers. Data by its nature is
confidential. When data is created by a human it may attract copyright
protection, and when created by an employee, that copyright vests in the
employer by virtue of the employment contract. A practical consequence
of confidentiality is that an organisation cannot readily transition
between a cost basis (D-Val) and a market-observed fair value without
exposing the dataset to a market that does not yet exist. In the absence
of a ready mechanism to elicit market value, the organisation defaults
to the cost basis. The A-Val construct provides an intermediate step: a
commercial valuation grounded in the characteristics of the dataset and
in theoretical premia derived from related certification markets,
defensible in negotiation, and suitable for internal decision-making,
without prematurely asserting auditable fair value.

This progression is not without market precedent. The Black-Scholes
(Black \& Scholes, 1973) formula, which underpins the derivatives market
(on conservative estimates exceeding US\$10 trillion in market value and
US\$1 quadrillion in notional value) was able to quantify the value of
an intangible asset. The A-Val construct performs an analogous function
for authenticated data assets.

\subsection{4.3 A-Val: The Commercial
Valuation}\label{a-val-the-commercial-valuation}

A-Val is the commercial valuation of an authenticated data asset. It
extends the D-Val cost basis by incorporating dataset-specific quality
attributes, market and legal context, and provisional premia for
authentication and audit verification. The A-Val formula is:

\[ A\text{-}Val = Cp \times \log_{10}(Sz)^{1.3} \times \frac{1}{e^{(Sc^{\beta})}} \times C \times Ac \times Pp \times \left( \frac{1}{No} \right) \times Ap \times Av^t \]

\paragraph{Cost floor}\label{cost-floor}

A-Val is bounded below by D-Val. Where the formula above produces a
value less than D-Val, the reported valuation is set to D-Val. Formally:

\emph{Reported value = max(D-Val, A-Val calculated)}

The cost floor is important because A-Val can fall below D-Val in cases
of severe scarcity erosion. Where an authenticated dataset is licensed
to multiple parties, the per-licensee commercial value may be low even
though the dataset itself, measured at cost, retains its replacement
value. The cost floor ensures the reported valuation never falls below
the auditable accounting floor. Section 5.3 illustrates this mechanism.

\subsubsection{4.3.1 Parameter Definitions}\label{parameter-definitions}

The A-Val parameters are summarised in Table 1 and defined in detail
thereafter.

{\def\LTcaptype{none}
\begin{longtable}[]{@{}
  >{\raggedright\arraybackslash}p{(\linewidth - 6\tabcolsep) * \real{0.3419}}
  >{\raggedright\arraybackslash}p{(\linewidth - 6\tabcolsep) * \real{0.1923}}
  >{\raggedright\arraybackslash}p{(\linewidth - 6\tabcolsep) * \real{0.2521}}
  >{\raggedright\arraybackslash}p{(\linewidth - 6\tabcolsep) * \real{0.2137}}@{}}
\toprule\noalign{}
\textbf{Name} & \textbf{Variable} & \textbf{Unit} & \textbf{Expected range} \\
\midrule\noalign{}
\endhead
\bottomrule\noalign{}
\endlastfoot
Commercial value of data & A-Val & \$ & A-Val $\geq$ D-Val \\
Cost to produce data & Cp & \$ & $0 \leq$ Cp \\
Appreciation or depreciation factor & Av & dimensionless & $0 <$ Av (constrained in D-Val) \\
Dataset size & Sz & Gigabytes & $1 \leq$ Sz \\
Scarcity & Sc & Count of non-licensor entities with access & $1 \leq$ Sc (1 = exclusive) \\
Completeness & C & dimensionless & $0 \leq$ C $\leq 1$ \\
Accuracy & Ac & dimensionless & $0 \leq$ Ac $\leq 1$ \\
Rivalry factor & $\beta$ & dimensionless & $0 \leq \beta \leq 1$ \\
Time & t & Years & $0 \leq$ t \\
Number of legal owners & No & Count & $1 \leq$ No \\
Provenance premium & Pp & dimensionless & Pp = 1.2 (provisional) \\
Audit premium & Ap & dimensionless & Ap = 1.2 (provisional) \\
\end{longtable}
}

\emph{Table 1. A-Val parameters.}

The parameters are defined as follows.

\begin{itemize}
\item
  \textbf{A-Val.} The commercial valuation of the authenticated data
  asset. Produced by the formula above and floored at D-Val.
\item
  \textbf{Cp.} The reliably measurable production cost of the dataset,
  consistent with IAS 38 cost measurement. The same Cp value enters both
  D-Val and A-Val.
\item
  \textbf{Av.} The appreciation or depreciation factor applied per unit
  time. Av greater than 1 indicates appreciation and less than 1
  indicates depreciation. Within D-Val, Av is constrained to values less
  than or equal to 1 absent an active market revaluation. Within A-Val,
  Av is unconstrained, reflecting that the commercial valuation is
  forward-looking and need not be limited by IAS 38 measurement
  restrictions.
\item
  \textbf{Sz.} Dataset size, in gigabytes (uncompressed). The
  logarithmic transformation captures the diminishing marginal value of
  additional volume. Size enters as a multiplicative factor on cost in
  A-Val because cost scales approximately with size but commercial value
  scales sub-linearly with size. \footnote{There are a few methods to calculating the size of data: 1) Size in bytes (uncompressed): The problem with this approach is that some modalities of data have very high redundancy, like videos or high-resolution images. 2) Size in bytes (compressed): The problem with this approach is that it depends on the compression scheme. 3) Number of data points: For each data modality, we choose a data point unit, for example, single words for Language and images for Vision. 4) Information-theoretic measures: such as Kolmogorov complexity. See: Pablo Villalobos and Anson Ho (2022), "Trends in training dataset sizes". Published online at epoch.ai. Retrieved from 'https://epoch.ai/blog/trends-in-training-dataset-sizes' [online resource]. Accessed 27 May 2026.}
\item
  \textbf{Sc.} The scarcity parameter, measured as the number of
  entities with access to the dataset. Sc = 1 denotes exclusive
  licensing. Higher values of Sc denote dilution. Valuation of
  non-exclusive licensing may function more effectively where licences
  are issued in capped quantities analogous to limited editions in art
  reproduction markets. A licensee may, for example, acquire licence
  number 4 of 30, thereby knowing that only a finite number of other
  parties will receive equivalent rights to the dataset.
\item
  \textbf{C.} Completeness, expressed as a fraction between 0 and 1. A
  value of 1 indicates a fully complete dataset; values below 1 reflect
  missing fields, missing records, or structural gaps.
\item
  \textbf{Ac.} Accuracy, expressed as a fraction between 0 and 1. A
  value of 1 indicates all records are correct; values below 1 reflect
  error rates observed in validation.
\item
  \textbf{$\beta.$} The rivalry factor, bounded between 0 (perfectly
  non-rivalrous) and 1 (perfectly rivalrous). Most digital datasets are
  technically non-rivalrous; contractual and regulatory context
  introduces effective rivalry. The value is determined systematically
  from data type, sensitivity, and regulatory context (see Section 5 for
  applied examples).
\item
  \textbf{No.} The number of legal owners of the asset. Where ownership
  is shared equally, the value attributable to a given owner is the
  reciprocal of No.
\item
  \textbf{Pp.} The provenance premium, set at 1.2 as a provisional
  parameter. Theoretical grounding is provided in Section 4.4.
\item
  \textbf{Ap.} The audit premium, set at 1.2 as a provisional parameter.
  Theoretical grounding is provided in Section 4.5.
\end{itemize}

\subsubsection{4.3.2 Limitation: Parameter
Backlogging}\label{limitation-parameter-backlogging}

A-Val is a forward-looking construct whose parameters require
calibration against transaction data that does not yet exist at scale.
Current applications of A-Val therefore rely on provisional parameter
values informed by related certification markets and by theoretical
argument. As authenticated data markets mature and transaction data
becomes available, valuers will be able to estimate the input parameters
from observed prices for datasets with similar characteristics. This
recursive relationship, in which each transaction refines the parameters
used to price the next, is analogous to the historical development of
option pricing methodology and is expected to produce progressively more
efficient price discovery. Section 6 identifies priority research
directions for this empirical programme.

The cost of production, dataset size, completeness, number of owners,
provenance premium, and audit premium are tangible metrics that can be
determined directly from empirical metadata without complex calculation.
In contrast, the scarcity, accuracy, rivalry, and appreciation factors
are latent variables that are difficult to attribute initial values to.

To estimate these latent parameters, we employ a two-stage supervised
learning approach using a corpus of historical dataset sales with known
transaction prices:

Using K-Means clustering, using a Euclidean distance to partition
historical datasets based on their observable features. For each
cluster, we derive initial estimates of the latent parameters.

Using the cluster-initialized parameters as starting values, we apply
constrained non-linear least squares regression to minimize the error
function. The regression optimizes the latent parameters across all
historical sales to achieve the best fit between the theoretical A-Val
formula and observed market prices. With the error function minimised,
the valuation formula produces a mathematically justified asset figure
that closely reflects the actual market value of the newly valued
dataset.

\subsection{\texorpdfstring{4.4 The Provenance Premium:
}{4.4 The Provenance Premium: }}\label{the-provenance-premium}

The A-Val formula incorporates an initial baseline 20\% value premium
(Pp = 1.2) for authenticated data assets with documented legal
provenance, verified ownership chains, and established usage rights.
This subsection presents the theoretical rationale supporting this
baseline parameter value and introduces in section 4.4.5 a variable and
bespoke value premium adjusted on an asset-by asset basis associated
with that asset's authentication ``metadata'' and ``secondary data''.

Across sports memorabilia, foodstuffs, luxury horology, and sustainable
bonds, authentication provides certainty in asset quality and
authenticity (Dupreele et al., 2023). In sports memorabilia,
authentication can lead to a 300\% price increase (Burk, 2025), while
certified organic foods command a 15 to 30\% premium (Tully and Winer,
2021). Green bonds trade 2 to 10 basis points lower than conventional
bonds (Panizza et al., 2025), demonstrating the tangible value of their
certified status. This evidence provides an empirical basis for the
provenance premium: the principle that authenticated assets command
higher valuations in the marketplace.

\subsubsection{4.4.1 Risk Reduction Through Information Asymmetry
Mitigation}\label{risk-reduction-through-information-asymmetry-mitigation}

Akerlof (1970) demonstrates that information asymmetry in markets with
quality uncertainty creates adverse selection problems, depressing asset
values below their true worth.\footnote{The central thesis in Akerlof's
  market for lemons, whilst subject to extensive commentary, has
  remained a reliable theorem with extensive support (Bar-Isaac, Jewitt
  and Leaver (2021)).} In the absence of credible quality signals,
buyers cannot distinguish high-quality assets from low-quality assets,
leading to market failure or severe undervaluation. Authentication
serves as a credible quality signal that mitigates this market for
lemons problem by providing verifiable evidence of legal, compliance,
and operational risk reduction.

\begin{itemize}
\item
  \textbf{Legal risk reduction.} Documented consent trails, ownership
  chains, and usage rights reduce litigation exposure under privacy
  regulations (GDPR, CCPA, HIPAA) and intellectual property law.
  Authenticated data enables buyers to verify compliance without
  extensive legal due diligence.
\item
  \textbf{Compliance risk reduction.} Verified provenance facilitates
  regulatory compliance demonstrations, reducing audit costs and the
  probability of regulatory sanction. Authentication documentation
  provides ready evidence for regulatory inquiries.
\item
  \textbf{Operational risk reduction.} Quality certification through
  authentication reduces data integration costs and downstream error
  correction expenses. Authenticated data comes with verified
  completeness and accuracy metrics.
\end{itemize}

Authenticated data provides asset holders with greater flexibility to
monetise assets through licensing or sale without triggering legal
challenges, equivalent to holding a real option on future
commercialisation. The premium reflects the value of this embedded
optionality.

\subsubsection{4.4.2 Transaction Cost Reduction Through Property Rights
Clarity}\label{transaction-cost-reduction-through-property-rights-clarity}

Coase (1960) establishes that clearly defined property rights reduce
transaction costs in asset markets.\footnote{Although the Coase Theorem
  was published over 60 years ago, and has faced criticism, there is
  still general support for the aspect elucidated being; clearly defined
  property rights reduce transaction costs in markets p (Medema (2020);
  Rindfleisch (2020)).} When property rights are ambiguous, parties must
invest substantial resources in establishing ownership, negotiating
terms, and enforcing agreements. Authentication establishes clear
property rights for data assets, reducing search and information costs,
bargaining and decision costs, and enforcement and policing costs.

\begin{itemize}
\item
  \textbf{Search and information costs.} Buyers can verify data quality,
  provenance, and legal status without extensive due diligence. Standard
  authentication frameworks provide common information infrastructure
  that reduces buyer investigation costs.
\item
  \textbf{Bargaining and decision costs.} Clear property rights reduce
  negotiation complexity. When ownership and usage rights are
  unambiguous, parties can focus on price rather than litigating
  fundamental rights questions.
\item
  \textbf{Enforcement and policing costs.} Documented rights facilitate
  contract enforcement and dispute resolution. Authentication provides
  clear evidence for breach of contract claims and intellectual property
  enforcement actions.
\end{itemize}

Barzel (1982) demonstrates that transaction costs can represent 20 to
40\% of asset value in markets with poorly defined property rights.
Authentication's reduction of these costs translates directly into
premium valuations, as buyers are willing to pay more for assets with
clear, enforceable rights.

\subsubsection{4.4.3 Network Effects and Market Liquidity
Enhancement}\label{network-effects-and-market-liquidity-enhancement}

Katz and Shapiro (1985) demonstrate that standardisation increases asset
liquidity by expanding the potential buyer pool and reducing
buyer-specific costs. As authentication frameworks become standardised
across data markets, authenticated assets benefit from positive network
externalities through standardised comparison frameworks, expanded
market participation, and enhanced liquidity.

\begin{itemize}
\item
  \textbf{Standardised comparison frameworks} enable buyers to compare
  authenticated datasets across different sellers using common metrics
  for provenance quality, completeness, and accuracy.
\item
  \textbf{Expanded market participation} results as reduced asymmetric
  information attracts more buyers and sellers to authenticated data
  markets.
\item
  \textbf{Enhanced liquidity} supports higher asset valuations (Amihud
  \& Mendelson, 1986). Assets that can be readily resold or relicensed
  command premium prices reflecting the value of liquidity optionality.
\end{itemize}

The liquidity premium component reflects buyers' willingness to pay more
for assets that can be readily resold or relicensed in established
markets, a consideration particularly important for data assets whose
value may evolve as market conditions and regulatory requirements
change.

\subsubsection{4.4.4 Parameter Specification and
Justification}\label{parameter-specification-and-justification}

Based on empirical evidence from related certification markets (15 to
40\% observed premia), we propose Pp = 1.2 as a provisional parameter
requiring validation. Future research should estimate this parameter
through regression analysis of authenticated versus non-authenticated
data transactions. Prior to such validation, valuers should treat the
parameter as a subjective assumption requiring sensitivity analysis and
potentially company-specific adjustment. This parameter reflects three
considerations.

\begin{itemize}
\item
  \textbf{Conservative economic positioning.} The 20\% premium
  represents a conservative estimate of combined risk reduction,
  transaction cost savings, and liquidity enhancement benefits. Barzel's
  (1982) finding that transaction costs represent 20 to 40\% of asset
  value in markets with unclear property rights suggests authentication
  could justify premia at the upper end of this range.
\item
  \textbf{Accounting conservatism alignment.} The parameter embodies
  conservative accounting principles by adopting a mid-point estimate
  rather than maximum theoretical values. This approach reduces
  professional risk for auditors and valuation specialists.
\item
  \textbf{Contextual adjustment framework.} While 20\% represents the
  base parameter, specific applications may warrant adjustment. Higher
  premia (1.4 to 1.5) may be appropriate for highly regulated sectors
  such as healthcare and finance, where compliance risk reduction
  provides substantial value. Lower premia (1.05 to 1.10) may apply to
  commoditised data types with minimal legal risk or established markets
  with transparent pricing.
\end{itemize}

Future research should examine authentication premium variation across
industries, jurisdictions, and data types as markets mature and
transaction data becomes more readily available. Regression analysis
relating authentication characteristics to transaction prices will
enable empirical validation and potential parameter refinement.

\subsubsection{4.4.5 The Provenance Premium: Accounting for Metadata and
Secondary
Data}\label{the-provenance-premium-accounting-for-metadata-and-secondary-data}

The authors anticipate that the initial baseline 20\% value premium (Pp
= 1.2) for authenticated data assets will evolve into a variable and
bespoke value premium adjusted on an asset-by asset basis. Metadata and
Secondary data are the labels given to describe those aspects of legal
authentication that are relevant to future bespoke provenance premium
values.

Metadata is data that describes the underlying dataset: its name, its
summary, and its identifiers, meaning things like the root asset ID and
this asset\textquotesingle s own repository access path. Secondary Data
consists of two subsets platform-use activity, and performance data of
the underlying container contracts. Both sets record things such as
movement of data, owners, access levels, and transaction history but the
performance data records the activity to do with the underlying contract
that governs the dataset.

Metadata and Secondary data both have practical worth to a dataset.
Knowing information about the dataset where it has been how it works and
who has access to the given dataset will have some effect on the
utility, trustworthiness and as a result the value of the underlying
data. As with the central data valuation problem, the process of
attributing a monetary figure to value this type of data is not
immediately evident.

To account for metadata in the valuation of a dataset, it is worth
noting that metadata is generally always available with the dataset
itself. Having access to the dataset it will be standard that one will
be provided with the date of creation size and its basic identifiers. As
it is standard for the metadata to be available, the value of having
access to this information will be accounted for in the underlying
dataset value.

Secondary data is more elusive and is only accessible with a concerted
effort to record the permissions access and activity of the platform
use. Because of this, a dataset with this information about the set will
have some marked increase in value separate to accessing the dataset by
itself. This increase in value will likely correspond to the total value
of the underlying dataset, instead of being calculable as a separate
dataset.

Secondary data will include things such as the owners, its permissioning
levels, a transcript of its history, and tracing of any changes to the
dataset. This data will be accounted for as the provenance premium.
Access to this information will allow the user to be confident in the
provenance of the given dataset.

\subsection{4.5 The Audit Premium: Theoretical
Foundations}\label{the-audit-premium-theoretical-foundations}

The A-Val formula incorporates a 20\% value premium (Ap = 1.2) for data
assets that have been independently audited and verified for quality,
completeness, and accuracy. This subsection presents the theoretical
rationale for this parameter.

\subsubsection{4.5.1 Certification Value and Signalling
Theory}\label{certification-value-and-signalling-theory}

Independent audit serves as a costly signal of data quality that
high-quality data asset owners can credibly send but low-quality asset
owners cannot economically replicate. Spence's (1973) signalling theory
demonstrates that in markets with information asymmetry, costly signals
can separate high-quality offerings from low-quality offerings, enabling
markets to function more efficiently. For data assets, three aspects of
signalling apply.

\begin{itemize}
\item
  \textbf{Signal credibility.} Independent third-party audits provide
  credible verification that self-reported quality metrics cannot match
  (Le et al., 2021). Buyers and their investors value audit verification
  because auditors face reputational and professional liability risks
  that discipline their assessments, reducing potential information
  asymmetries.
\item
  \textbf{Separating equilibrium.} Audit costs create a separating
  equilibrium where only data asset owners confident in their quality
  will submit to independent verification. This selection effect means
  audited data commands premium valuations reflecting both verified
  quality and revealed owner confidence.
\item
  \textbf{Repeated game dynamics.} Auditors who consistently verify
  high-quality data build reputational capital (separate from, and in
  addition to, first-mover advantage), enhancing the signal value of
  their certifications. Markets develop hierarchies of audit quality
  that buyers use to differentiate data assets.
\end{itemize}

The signalling value of audit verification reduces buyer uncertainty and
information asymmetry costs, translating directly into willingness to
pay premium prices for audited data assets.

\subsubsection{4.5.2 Agency Cost Reduction}\label{agency-cost-reduction}

Agency theory (Jensen \& Meckling, 1976) identifies costs arising from
information asymmetry between principals and agents. In data asset
transactions, sellers (agents) possess superior information about data
quality relative to buyers (principals), creating agency costs in the
form of monitoring costs, bonding costs, and residual loss.

\begin{itemize}
\item
  \textbf{Monitoring costs.} Without independent audit, buyers must
  invest substantial resources in verification activities including
  sample testing, quality assessment, and due diligence investigations.
  Independent audit transfers the monitoring function to specialised
  professionals who achieve economies of scale and expertise.
\item
  \textbf{Bonding costs.} Sellers submitting to independent audit
  effectively bond their quality representations, accepting professional
  scrutiny that would reveal misrepresentation. This bonding reduces
  buyer risk and associated risk premia.
\item
  \textbf{Residual loss reduction.} Even with monitoring and bonding,
  information asymmetry creates residual losses from suboptimal
  decisions. Independent audit reduces residual loss by providing
  objective quality assessments that enable more accurate pricing and
  utilisation decisions.
\end{itemize}

The reduction in agency costs through independent audit justifies
premium valuations, as these cost savings accrue to both buyers (reduced
due diligence costs) and sellers (reduced cost of capital through
enhanced credibility) (Ullah, 2020).

\subsubsection{4.5.3 Insurance Value and Risk
Transfer}\label{insurance-value-and-risk-transfer}

Independent audit provides implicit insurance against data quality
defects. Professional auditors carry liability insurance and face
reputational consequences for verification failures, effectively
transferring some quality risk from data buyers to audit firms through
three mechanisms.

\begin{itemize}
\item
  \textbf{Professional liability.} Audit firms face professional
  liability exposure for negligent verification. This liability creates
  incentives for thorough quality assessment and provides buyers with
  recourse if audited data proves defective.
\item
  \textbf{Reputational capital at risk.} Established audit firms possess
  significant reputational capital that they risk through poor quality
  verification. This capital serves as implicit collateral backing audit
  quality.
\item
  \textbf{Risk pooling.} Audit firms pool risks across multiple clients
  and engagements, achieving risk diversification that individual data
  buyers cannot replicate. This risk pooling enables more efficient risk
  bearing.
\end{itemize}

The insurance value of audit verification represents a real economic
benefit that justifies premium asset valuations, as buyers effectively
purchase both the data asset and partial insurance against quality
defects.

\subsubsection{4.5.4 Empirical Evidence from Related
Markets}\label{empirical-evidence-from-related-markets}

While comprehensive empirical studies of data asset audit premia remain
limited due to market nascency, evidence from related certification and
audit markets suggests substantial value premia.

Financial statement audits. Companies with audited financial statements
experience measurably lower costs of capital than comparable unaudited
firms (Coffie et al., n.d.). Research in accounting economics documents
that audit quality (often proxied by auditor size or specialisation)
reduces information risk premia in equity markets by 20 to 40 basis
points in required returns. This cost of capital reduction translates to
substantial value increases, supporting the premise that independent
verification commands premium valuations.

Quality management system certification. ISO 9001 and similar quality
certifications require independent audit verification. Manufacturing
firms with ISO certification report price premia of 15 to 35\% (Levine
and Toffel, 2008) in business-to-business transactions relative to
uncertified competitors, reflecting buyer willingness to pay for
verified quality assurance.

Environmental and social certifications. Products with independently
verified environmental or social certifications (carbon neutral, fair
trade, sustainable sourcing) command price premia averaging 20 to 40\%
across consumer and industrial markets. These premia reflect consumer
and buyer willingness to pay for credible third-party verification
(Panizza et al., 2025).

\subsubsection{4.5.5 Parameter Specification and
Justification}\label{parameter-specification-and-justification-1}

Based on theoretical frameworks and evidence from related certification
markets, we specify Ap = 1.2 (20\% premium) for independently audited
data assets. This parameter reflects three considerations.

\begin{itemize}
\item
  \textbf{Conservative estimate.} The 20\% premium represents a
  conservative estimate relative to observed premia in quality
  certification markets (15 to 40\%), positioning the parameter at the
  lower-middle range of comparable certification values.
\item
  \textbf{Multiplicative structure justification.} The audit premium (Ap
  = 1.2) applies multiplicatively with the provenance premium (Pp =
  1.2), yielding a combined 44\% premium (1.2 × 1.2 = 1.44) for data
  assets that are both authenticated and audited. This multiplicative
  structure reflects that provenance authentication and quality audit
  address different information asymmetries: provenance verifies legal
  rights and chain of custody, while audit verifies data quality and
  accuracy.
\item
  \textbf{Context-dependent adjustment.} The 20\% base parameter may
  warrant adjustment based on audit quality and scope. Premium audits by
  leading specialised firms covering comprehensive quality dimensions
  may justify higher premia (1.4 to 1.5), while basic quality checks may
  support lower premia (1.05 to 1.1).
\end{itemize}

As data asset markets mature and specialised data quality audit
practices develop, empirical research should examine audit premium
variation across audit firm types, audit scope specifications, and data
asset characteristics. Such research will enable evidence-based
parameter refinement and potentially dynamic premium calculation based
on observable audit quality indicators.

\subsubsection{4.5.6 Combined Authentication and Audit
Effects}\label{combined-authentication-and-audit-effects}

The A-Val formula's multiplicative structure for the provenance premium
(Pp) and the audit premium (Ap) deserves explicit justification. The
combined 44\% premium (1.2 × 1.2 = 1.44) for authenticated and audited
data assets reflects complementary rather than redundant value creation.

\begin{itemize}
\item
  \textbf{Distinct information asymmetries.} Authentication addresses
  legal and provenance uncertainty (ownership rights, usage permissions,
  chain of custody), while audit addresses quality uncertainty
  (completeness, accuracy, consistency). These represent different
  dimensions of information asymmetry that buyers value independently.
\item
  \textbf{Complementary risk reduction.} Legal risk reduction through
  authentication and quality risk reduction through audit provide
  complementary benefits. A data asset with clear legal rights but poor
  quality, or high quality but unclear rights, remains problematic.
  Combined authentication and audit address both risk dimensions
  comprehensively.
\item
  \textbf{Empirical support.} Markets demonstrate that combined
  certifications command premia exceeding individual certification
  effects. Products with both provenance certification (geographic
  indication, organic sourcing) and quality certification (grading,
  testing) command premia substantially higher than either certification
  alone.
\end{itemize}

The combined 44\% premium for authenticated and audited data assets thus
represents a conservative estimate of value creation through
comprehensive information asymmetry reduction across legal and quality
dimensions.

\subsection{4.6 Parameter Uncertainty and
Sensitivity}\label{parameter-uncertainty-and-sensitivity}

The A-Val formula's multiplicative structure amplifies parameter
uncertainty. We identify high-impact parameters requiring careful
estimation.

\begin{itemize}
\item
  \textbf{Critical parameters ($>$20\% valuation impact per
  10\% change).} The rivalry factor ($\beta$) directly moderates scarcity
  effects; $\beta = 0.6$ versus $\beta = 0.7$ creates 12 to 15\% valuation
  differences in shared datasets. The authentication and audit premia
  (Pp, Ap) have a combined multiplicative effect yielding roughly 21\%
  impact per 10\% parameter change.
\item
  \textbf{Moderate parameters (10 to 20\% impact).} Completeness and
  accuracy (C, Ac): a combined ten-percentage-point reduction creates
  approximately a 15\% valuation decrease. The appreciation factor
  ($Av^t$): impact increases with time horizon.
\item
  \textbf{Low-impact parameters ($<$10\% impact).} Size factor
  (Sz): logarithmic dampening limits sensitivity. Ownership division
  (1/No): typically binary.
\end{itemize}

Practitioners should prioritise estimation accuracy for $\beta$, Pp, and Ap.
Where material uncertainty exists, range reporting provides appropriate
disclosure. Conservative, central, and optimistic estimates should be
constructed as follows: conservative estimate uses $\beta + 0.1$, $P_p = A_p =
1.1$, quality metrics reduced by 5\%; central estimate uses framework-determined values; 
optimistic estimate uses $\beta - 0.1$, $P_p = A_p = 1.4$, quality metrics increased by 5\%.

\section{5. Worked Examples}\label{worked-examples}

This section presents three detailed examples demonstrating the
two-layer progression across distinct industry contexts: retail customer
analytics, mining geological data, and healthcare clinical outcomes.
Each example reports D-Val (the auditable cost-basis valuation), A-Val
central, conservative, and optimistic estimates, and market validation
against available comparables.

\subsection{5.1 Example 1: Retail Customer Transaction
Database}\label{example-1-retail-customer-transaction-database}

A mid-sized retail company developed a comprehensive customer
transaction database over five years, containing purchase histories,
product preferences, demographic information, and behavioural analytics
for 2.5 million customers. The company seeks to value this asset for
potential licensing to non-competing businesses in complementary
sectors.

\paragraph{Dataset characteristics}\label{dataset-characteristics}

{\def\LTcaptype{none} 
\begin{longtable}[]{@{}
  >{\raggedright\arraybackslash}p{(\linewidth - 2\tabcolsep) * \real{0.3205}}
  >{\raggedright\arraybackslash}p{(\linewidth - 2\tabcolsep) * \real{0.6795}}@{}}
\toprule\noalign{}
\endhead
\bottomrule\noalign{}
\endlastfoot
Production cost (Cp) & \$750,000 (data scientists, infrastructure, five
years' maintenance) \\
Size (Sz) & 250 GB \\
Scarcity (Sc) & 1 (exclusive) \\
Completeness (C) & 0.94 (some missing demographic fields) \\
Accuracy (Ac) & 0.91 (validated against known purchases) \\
Authentication & Yes (complete set of contractual artefacts and legal
registration) \\
Audit status & Audited by external data quality firm \\
Appreciation (Av) & 1.08 annually (growing customer base) \\
Age (t) & 2 years since major overhaul \\
\end{longtable}
}

\paragraph{D-Val calculation}\label{d-val-calculation}

Under IAS 38 as currently applied, Av is constrained to 1 in the absence
of an active market revaluation. For the D-Val calculation:

D-Val = Cp × Av\^{}t = \$750,000 × 1.0² = \$750,000

The auditable carrying value is \$750,000 less any accumulated
amortisation. For the purposes of this example, amortisation is assumed
to be nil, and D-Val = \$750,000.

\paragraph{Determination of β}\label{determination-of-ux3b2}

\begin{itemize}
\item
  Data type: commercial (base $\beta$ = 0.5).
\item
  Sensitivity: medium (+0.1); moderate competitive sensitivity.
\item
  Regulatory: no additional adjustment (+0.0); consumer privacy laws
  apply but not HIPAA or financial services regulation.
\item
  Calculated $\beta$ = 0.6.
\end{itemize}

\paragraph{A-Val calculation}\label{a-val-calculation}

{\def\LTcaptype{none} 
\begin{longtable}[]{@{}
  >{\raggedright\arraybackslash}p{(\linewidth - 2\tabcolsep) * \real{0.5342}}
  >{\raggedright\arraybackslash}p{(\linewidth - 2\tabcolsep) * \real{0.4658}}@{}}
\toprule\noalign{}
\endhead
\bottomrule\noalign{}
\endlastfoot
Size factor: $\log_{10}(250)^{1.3}$ & 3.117 \\
Scarcity factor: $1 / e^{(1^{0.6})}$ & 0.368 \\
Completeness (C) & 0.94 \\
Accuracy (Ac) & 0.91 \\
Provenance premium (Pp) & 1.2 \\
Ownership (1/No) & 1.0 \\
Audit premium (Ap) & 1.2 \\
Time factor ($Av^t$): $1.08^2$ & 1.166 \\
\end{longtable}
}

A-Val = \$750,000 × 3.117 × 0.368 × 0.94 × 0.91 × 1.2 × 1.0 × 1.2 ×
1.166

A-Val = \$1,235,765 (central estimate).

\paragraph{Valuation summary}\label{valuation-summary}

{\def\LTcaptype{none} 
\begin{longtable}[]{@{}
  >{\raggedright\arraybackslash}p{(\linewidth - 2\tabcolsep) * \real{0.4915}}
  >{\raggedright\arraybackslash}p{(\linewidth - 2\tabcolsep) * \real{0.5085}}@{}}
\toprule\noalign{}
\endhead
\bottomrule\noalign{}
\endlastfoot
D-Val (auditable cost basis) & \$750,000 \\
A-Val central estimate & \$1,235,765 \\
A-Val as multiple of D-Val & 1.65x \\
Commercial uplift over cost basis & \$485,765 (65\%) \\
\end{longtable}
}

\paragraph{Sensitivity analysis}\label{sensitivity-analysis}

Applying the sensitivity framework from Section 4.6:

{\def\LTcaptype{none} 
\begin{longtable}[]{@{}
  >{\raggedright\arraybackslash}p{(\linewidth - 2\tabcolsep) * \real{0.6197}}
  >{\raggedright\arraybackslash}p{(\linewidth - 2\tabcolsep) * \real{0.3803}}@{}}
\toprule\noalign{}
\endhead
\bottomrule\noalign{}
\endlastfoot
Conservative A-Val ($\beta$ = 0.7, Pp = Ap = 1.1, quality −5\%) &
approximately \$929,000 \\
Central A-Val & \$1,235,765 \\
Optimistic A-Val ($\beta$ = 0.5, Pp = Ap = 1.4, quality +5\%) & approximately
\$1,869,000 \\
\end{longtable}
}

All three scenarios produce A-Val estimates above D-Val of \$750,000, so
the cost floor does not activate in this example. Where a conservative
A-Val scenario did fall below D-Val, the cost floor provision would set
the reported conservative value to D-Val, ensuring the commercial
valuation cannot pull the reported number below the auditable accounting
floor. The healthcare example in Section 5.3 illustrates this mechanism.

\paragraph{Market validation}\label{market-validation}

Comparable retail customer database transactions include: a similar
grocery chain dataset (3 million customers) licensed at approximately
\$1.2M annually; UK retail consortium individual firm stakes valued at
£600K to £900K; and US credit bureau customer segments at \$0.30 to
\$0.50 per record. The central A-Val of \$1,235,765 sits within the
range suggested by these comparables. Annual licensing at 15 to 20\% of
asset value (\$185K to \$247K) aligns with observed market rates.

\subsection{5.2 Example 2: Mining Geological Survey
Dataset}\label{example-2-mining-geological-survey-dataset}

A mining exploration company conducted extensive geological surveys
across 5,000 square kilometres in Western Australia, including seismic
data, core sample analyses, geochemical assays, and 3D geological models
developed over eight years.

\paragraph{Dataset characteristics}\label{dataset-characteristics-1}

{\def\LTcaptype{none} 
\begin{longtable}[]{@{}
  >{\raggedright\arraybackslash}p{(\linewidth - 2\tabcolsep) * \real{0.3205}}
  >{\raggedright\arraybackslash}p{(\linewidth - 2\tabcolsep) * \real{0.6795}}@{}}
\toprule\noalign{}
\endhead
\bottomrule\noalign{}
\endlastfoot
Production cost (Cp) & \$8,500,000 \\
Size (Sz) & 12 TB (12,000 GB) \\
Scarcity (Sc) & 1 (exclusive) \\
Completeness (C) & 0.97 \\
Accuracy (Ac) & 0.93 \\
Authentication & Yes \\
Audit status & Audited \\
Depreciation (Av) & 0.96 annually \\
Age (t) & 4 years \\
\end{longtable}
}

\paragraph{D-Val calculation}\label{d-val-calculation-1}

The dataset is depreciating, so Av \textless{} 1 and D-Val reflects
depreciation:

\(\text{D-Val} = C_p \times \text{Av}^t = \$8,500,000 \times 0.96^4 = \$8,500,000 \times 0.849 = \$7,219,446.\)

The auditable carrying value is \$7,219,446 (cost less the cumulative
depreciation factor).

\paragraph{Determination of β}\label{determination-of-ux3b2-1}

\begin{itemize}
\item
  Data type: proprietary (base $\beta$ = 0.7).
\item
  Sensitivity: high (+0.2).
\item
  Regulatory: no additional adjustment (+0.0).
\item
  Calculated $\beta$ = 0.9.
\end{itemize}

\paragraph{A-Val calculation}\label{a-val-calculation-1}

{\def\LTcaptype{none} 
\begin{longtable}[]{@{}
  >{\raggedright\arraybackslash}p{(\linewidth - 2\tabcolsep) * \real{0.5342}}
  >{\raggedright\arraybackslash}p{(\linewidth - 2\tabcolsep) * \real{0.4658}}@{}}
\toprule\noalign{}
\endhead
\bottomrule\noalign{}
\endlastfoot
Size factor: $\log_{10}(12,000)^{1.3}$ & 6.219 \\
Scarcity factor: $1 / e^{(1^{0.9})}$ & 0.368 \\
Completeness (C) & 0.97 \\
Accuracy (Ac) & 0.93 \\
Provenance premium (Pp) & 1.2 \\
Ownership (1/No) & 1.0 \\
Audit premium (Ap) & 1.2 \\
Time factor ($Av^t$): $0.96^4$ & 0.849 \\
\end{longtable}
}

A-Val = \$8,500,000 × 6.219 × 0.368 × 0.97 × 0.93 × 1.2 × 1.0 × 1.2 ×
0.849

A-Val = \$21,457,121 (central estimate).

\paragraph{Valuation summary}\label{valuation-summary-1}

{\def\LTcaptype{none} 
\begin{longtable}[]{@{}
  >{\raggedright\arraybackslash}p{(\linewidth - 2\tabcolsep) * \real{0.4915}}
  >{\raggedright\arraybackslash}p{(\linewidth - 2\tabcolsep) * \real{0.5085}}@{}}
\toprule\noalign{}
\endhead
\bottomrule\noalign{}
\endlastfoot
D-Val (auditable cost basis) & \$7,219,446 \\
A-Val central estimate & \$21,457,121 \\
A-Val as multiple of D-Val & 2.97x \\
A-Val as multiple of Cp & 2.52x \\
\end{longtable}
}

\paragraph{Sensitivity analysis}\label{sensitivity-analysis-1}

{\def\LTcaptype{none} 
\begin{longtable}[]{@{}
  >{\raggedright\arraybackslash}p{(\linewidth - 2\tabcolsep) * \real{0.6197}}
  >{\raggedright\arraybackslash}p{(\linewidth - 2\tabcolsep) * \real{0.3803}}@{}}
\toprule\noalign{}
\endhead
\bottomrule\noalign{}
\endlastfoot
Conservative A-Val & approximately \$16,181,000 \\
Central A-Val & \$21,457,121 \\
Optimistic A-Val & approximately \$31,727,000 \\
\end{longtable}
}

\paragraph{Market validation}\label{market-validation-1}

Seismic data packages in offshore petroleum command US\$2M to US\$15M
per survey area. Regional geological datasets in Western Australia trade
at A\$5M to A\$12M for similar coverage. The central A-Val multiple of
2.97x D-Val reflects authenticated provenance providing IP protection,
and strategic value in active mining regions. Annual licensing at 10 to
12\% of A-Val would yield \$2.15M to \$2.58M, broadly consistent with
industry rates.

\subsection{5.3 Example 3: De-identified Clinical Outcomes
Database}\label{example-3-de-identified-clinical-outcomes-database}

A hospital network compiled clinical outcomes data covering 750,000
patient encounters over ten years, professionally de-identified to HIPAA
standards. The dataset includes treatment protocols, outcomes,
complications, and long-term follow-up across multiple therapeutic
areas.

\paragraph{Dataset characteristics}\label{dataset-characteristics-2}

{\def\LTcaptype{none} 
\begin{longtable}[]{@{}
  >{\raggedright\arraybackslash}p{(\linewidth - 2\tabcolsep) * \real{0.3205}}
  >{\raggedright\arraybackslash}p{(\linewidth - 2\tabcolsep) * \real{0.6795}}@{}}
\toprule\noalign{}
\endhead
\bottomrule\noalign{}
\endlastfoot
Production cost (Cp) & \$3,200,000 \\
Size (Sz) & 450 GB \\
Scarcity (Sc) & 3 (licensed to three pharma companies) \\
Completeness (C) & 0.89 \\
Accuracy (Ac) & 0.95 \\
Authentication & Yes \\
Audit status & Audited \\
Appreciation (Av) & 1.06 annually \\
Age (t) & 3 years \\
\end{longtable}
}

\paragraph{D-Val calculation}\label{d-val-calculation-2}

Under IAS 38 as currently applied, Av is constrained to 1 for D-Val
absent an active market revaluation:

D-Val = Cp × Av\^{}t = \$3,200,000 × 1.0³ = \$3,200,000.

The auditable carrying value is \$3,200,000 (no amortisation assumed for
this example).

\paragraph{Determination of β}\label{determination-of-ux3b2-2}

\begin{itemize}
\item
  Data type: personal (base $\beta$ = 0.8).
\item
  Sensitivity: critical (+0.1).
\item
  Regulatory: regulated (+0.1).
\item
  Calculated $\beta$ = 1.0 (perfect rivalry).
\end{itemize}

\paragraph{A-Val calculation}\label{a-val-calculation-2}

{\def\LTcaptype{none} 
\begin{longtable}[]{@{}
  >{\raggedright\arraybackslash}p{(\linewidth - 2\tabcolsep) * \real{0.5342}}
  >{\raggedright\arraybackslash}p{(\linewidth - 2\tabcolsep) * \real{0.4658}}@{}}
\toprule\noalign{}
\endhead
\bottomrule\noalign{}
\endlastfoot
Size factor: $\log_{10}(450)^{1.3}$ & 3.556 \\
Scarcity factor: $1 / e^{(3^{1.0})}$ & 0.050 \\
Completeness (C) & 0.89 \\
Accuracy (Ac) & 0.95 \\
Provenance premium (Pp) & 1.2 \\
Ownership (1/No) & 1.0 \\
Audit premium (Ap) & 1.2 \\
Time factor ($Av^t$): $1.06^3$ & 1.191 \\
\end{longtable}
}

A-Val (calculated) = \$3,200,000 × 3.556 × 0.050 × 0.89 × 0.95 × 1.2 ×
1.0 × 1.2 × 1.191

A-Val (calculated) = \$821,417.

Cost floor applied: reported A-Val = max(D-Val, A-Val calculated) =
max(\$3,200,000, \$821,417) = \$3,200,000.

\paragraph{Valuation summary}\label{valuation-summary-2}

{\def\LTcaptype{none} 
\begin{longtable}[]{@{}
  >{\raggedright\arraybackslash}p{(\linewidth - 2\tabcolsep) * \real{0.4915}}
  >{\raggedright\arraybackslash}p{(\linewidth - 2\tabcolsep) * \real{0.5085}}@{}}
\toprule\noalign{}
\endhead
\bottomrule\noalign{}
\endlastfoot
D-Val (auditable cost basis) & \$3,200,000 \\
A-Val calculated (pre-floor) & \$821,417 \\
A-Val reported (post-floor) & \$3,200,000 \\
A-Val as multiple of D-Val & 1.00x (floored) \\
\end{longtable}
}

\paragraph{Market validation}\label{market-validation-2}

Real-world evidence databases are routinely licensed at \$500K to \$2M
per indication. Claims databases (IQVIA, Optum) command \$1M to \$5M
annually for major pharma access. The reported valuation of \$3.2M total
(approximately \$1.07M per licensee if divided equally) aligns with
market expectations for shared clinical datasets. Annual licensing at 12
to 15\% of total value would yield \$384K to \$480K, or approximately
\$128K to \$160K per pharmaceutical company, consistent with
willingness-to-pay for non-exclusive real-world evidence.

\paragraph{Interpretation}\label{interpretation}

The calculated A-Val of \$821,417 falls substantially below the cost
basis, triggering the cost floor provision. This outcome is economically
sensible. The dataset exhibits perfect rivalry ($\beta$ = 1.0) and is shared
among three licensees (Sc = 3), substantially reducing commercial value
per entity. The scarcity factor of 0.050 reflects that each licensee's
value is severely diminished by competitors' access to identical data.

The cost floor ensures conservative valuation consistent with IAS 38's
prudence principle. The hospital network retains the asset at cost,
reflecting that while commercial value per entity is low due to sharing,
the dataset required \$3.2M to create and retains that replacement
value. The reported valuation is therefore D-Val.

This example illustrates an important property of the two-layer
progression. A-Val can fall below D-Val where market structure (in this
case, multi-licensee sharing under perfect rivalry) compresses
per-entity commercial value below the cost basis. The cost floor
prevents the commercial model from pulling the reported valuation below
the auditable floor. A-Val in this case signals that the licensing
structure has commercially suboptimal characteristics, while D-Val
continues to anchor the balance sheet at the economically defensible
replacement value.

\subsection{5.4 Cross-Example Analysis}\label{cross-example-analysis}

The three examples demonstrate how the two-layer progression responds to
different data characteristics and reveal several patterns.

\begin{itemize}
\item
  \textbf{D-Val is invariant to market structure.} In all three
  examples, D-Val is determined by production cost and the applicable
  amortisation or depreciation factor. It does not depend on scarcity,
  rivalry, or the number of licensees. This is the correct behaviour for
  an auditable cost-basis valuation and is consistent with IAS 38.
\item
  \textbf{A-Val responds to market structure, including in the downward
  direction.} The healthcare example illustrates that A-Val can fall
  below D-Val when multi-licensee sharing under perfect rivalry
  compresses commercial value per entity. The cost floor ensures the
  reported valuation never falls below D-Val.
\item
  \textbf{Rivalry and scarcity interact multiplicatively.} The
  healthcare example shows that high rivalry ($\beta$ = 1.0) combined with
  shared access (Sc = 3) creates severe value degradation per entity.
  The scarcity factor drops from 0.368 (exclusive access) to 0.050
  (three-way sharing), a reduction of 86\%.
\item
  \textbf{Size exhibits logarithmic returns.} The mining dataset is 48
  times larger than the retail dataset (12 TB versus 250 GB), but the
  size factor is only 2.0 times higher (6.219 versus 3.117), reflecting
  diminishing marginal value of additional data. This prevents
  unrealistic valuations for very large datasets.
\item
  \textbf{Quality metrics have moderate impact.} The healthcare
  dataset's lower completeness (0.89) versus the mining dataset's higher
  completeness (0.97) produces approximately an 8\% valuation difference
  in isolation. Combined with accuracy differences, quality factors
  account for roughly 10 to 15\% of total A-Val variance across the
  examples.
\item
  \textbf{Authentication and audit premia are substantial.} The combined
  44\% premium (1.2 × 1.2 = 1.44) applies uniformly across examples,
  reflecting that provenance and quality verification provide consistent
  value regardless of data type or industry context.
\item
  \textbf{A-Val multiples over D-Val span a reasonable range.} The
  observed central A-Val multiples (1.00x to 2.97x of D-Val) align with
  intangible asset valuation literature suggesting internally developed
  intangibles typically trade at 1.5x to 4.0x development cost when
  market values exceed book values. The lower end of this range (the
  healthcare case at 1.00x) reflects the floor operating as intended.
\end{itemize}

These patterns validate the formula's economic logic while highlighting
that parameter selection, particularly $\beta$ and Sc, substantially affects
A-Val outcomes. This emphasises the importance of careful rivalry
assessment, transparent documentation of assumptions, and comprehensive
sensitivity analysis in practical applications. It also emphasises the
value of reporting both D-Val and A-Val together, because the gap
between them carries interpretive weight: a large positive gap reflects
commercial value creation through authentication and audit, while a
floored A-Val signals that the licensing structure is suboptimal
relative to the dataset's replacement value.

\section{6. Conclusion and Implementation
Guidance}\label{conclusion-and-implementation-guidance}

This paper proposes a two-layer valuation progression for authenticated
data assets meeting IAS 38 recognition criteria. The first layer, D-Val,
is the auditable cost-basis valuation consistent with IAS 38 as
currently applied: $\text{D-Val} = Cp \times Av^t$, with Av constrained to values
less than or equal to 1 absent an active market revaluation. D-Val can
be placed on a balance sheet today, subject to the standard recognition
tests being met. The second layer, A-Val, is a theoretically grounded
commercial valuation that incorporates dataset-specific quality
attributes, scarcity, rivalry, and explicit premia for authentication
and audit verification. A-Val is bounded below by D-Val through a cost
floor provision, and serves as a defensible commercial estimate during
the period before active markets for authenticated data assets mature.

The contribution of this paper is threefold. First, it articulates D-Val
as the auditable number that can be recognised under IAS 38 today,
addressing the gap created by conservative application of the standard
to an emerging asset class. Second, it introduces A-Val as the
commercial bridge between the current cost-basis state and the future
fair-value state, with theoretical grounding in property rights theory,
signalling theory, and agency cost frameworks. Third, it identifies the
market-maturation pathway by which A-Val becomes a candidate for
auditable fair value measurement under the IAS 38 revaluation model,
once an active market for authenticated data assets exists.

The framework incorporates theoretically grounded premia for
authentication (20\%) and audit verification (20\%) based on property
rights theory, signalling theory, and observed premia in related
certification markets. While these parameters lack direct empirical
validation from authenticated data transactions, authenticated data
markets still being nascent, they are within the range observed in
non-authenticated certification transactions and provide defensible
starting points enabling practical application while awaiting market
maturation.

Three worked examples demonstrate methodology application across retail,
mining, and healthcare sectors, with A-Val central multiples ranging
from 1.00x (floored at D-Val) to 2.97x of D-Val. Market validation
against available comparables suggests reasonable alignment, though
comparable heterogeneity and market opacity limit definitive
confirmation.

\subsection{6.1 Limitations and Future
Research}\label{limitations-and-future-research}

Several limitations warrant acknowledgment. First, functional form
specifications (logarithmic size factor, exponential scarcity function)
represent reasonable choices lacking formal derivation. Alternative
specifications may prove superior as empirical data enables comparative
testing. Second, $\beta$ determination frameworks, while systematic, require
subjective judgment. Inter-rater reliability testing would strengthen
the approach. Third, the worked examples use sparse, heterogeneous
comparables; comprehensive validation requires denser transaction data
with standardised reporting. Fourth, the paper does not and cannot
determine the accounting treatment of any specific data asset;
recognition decisions must be made by the reporting entity and its
auditors with reference to the specific facts and the relevant
standards. Fifth, the paper does not address transfer pricing, tax, or
regulatory capital implications of authenticated data recognition; these
are important areas for future work.

As authenticated data markets develop, priority research should focus on
the construction of transaction databases enabling regression analysis
of authentication and audit effects on prices; functional form testing
comparing A-Val predictions against actual transactions;
industry-specific parameter estimation examining premium variation
across sectors; and dynamic modelling investigating how authentication
value evolves with market maturation.

Another area that could justify continued research and development on the model is the effect of entropy on data valuation. Specifically, we suggest it could be incorporated into the Sz variable which is currently assessed on an uncompressed basis. If the Sz variable was on a compressed basis and determinable on a standardised basis between data categories, we suggest this link as compression algorithms, more specifically maximum theoretical compression rates rely on entropy calculations. High entropy data tends to be much more valuable, as it is a measure of information within a set, which would generally correlate to usable information. That being said, we acknowledge that a random set also has high entropy but low usable information.

Organisations implementing this methodology should recognise
participation in market formation. Early valuations and subsequent
transactions generate empirical data enabling systematic refinement, an
inherent characteristic of emerging asset class development rather than
a methodological weakness.

Particular attention should be directed toward empirically validating
the authentication and audit premia across different industries and
regulatory environments, enabling potential development of
industry-specific premium parameters. As data markets mature and
transaction transparency increases, systematic refinement of formula
parameters will enhance valuation accuracy and professional acceptance.

\newpage

\section{References}\label{references}

Akerlof, G. A. (1970). The market for ``lemons'': Quality uncertainty
and the market mechanism. Quarterly Journal of Economics, 84(3),
488--500.

Amihud, Y., \& Mendelson, H. (1986). Asset pricing and the bid-ask
spread. Journal of Financial Economics, 17(2), 223--249.

Bar-Isaac, H., Jewitt, I., \& Leaver, C. (2021). Adverse selection,
efficiency and the structure of information. \emph{Economic Theory},
\emph{72}(2), 579--614.
https://doi.org/10.1007/s00199-020-01300-1Barzel, Y. (1982). Measurement
cost and the organisation of markets. Journal of Law and Economics,
25(1), 27--48.

Black, F., \& Scholes, M. (1973). The pricing of options and corporate
liabilities. Journal of Political Economy, 81(3), 637--654.

Burk, D. (2025). Sports memorabilia and autographs: Forgeries, the
authentication process, and the value of authenticity. Oregon State
University.

Coase, R. H. (1960). The problem of social cost. Journal of Law and
Economics, 3, 1--44.

Coffie, W., Bedi, I., \& Amidu, M. (n.d.). The effects of audit quality
on the costs of capital of firms in Ghana. Peer-reviewed article,
University of Ghana Business School.

Coyle, D., \& Manley, A. (2020). What is the value of data? A review of
empirical methods. National Institute Economic Review, 253, R15--R25.

Dupreelle, P., Willersdorf, S., Llinas, N., Schuler, M., \& Brennan, J.
(2023). Luxury preowned watches, your time has come. Boston Consulting
Group.

International Accounting Standards Board. (2018). IAS 38 Intangible
Assets. IFRS Foundation.

International Accounting Standards Board. (2018). IFRS 3 Business
Combinations. IFRS Foundation.

Jensen, M. C., \& Meckling, W. H. (1976). Theory of the firm: Managerial
behavior, agency costs and ownership structure. Journal of Financial
Economics, 3(4), 305--360.

Katz, M. L., \& Shapiro, C. (1985). Network externalities, competition,
and compatibility. American Economic Review, 75(3), 424--440.

Le, H. T. T., Tran, H. G., \& Vo, X. V. (2021). Audit quality, accruals
quality and the cost of equity in an emerging market: Evidence from
Vietnam. International Review of Financial Analysis, 77, 101798.

Levine, D. I., \& Toffel, M. W. (2008). Quality management and job
quality: How the ISO 9001 standard for quality management systems
affects employees and employers. IRLE Working Paper No. 172-08,
University of California, Berkeley.

Medema, Steven G. \emph{The Coase Theorem at Sixty} Journal of Economic
Literature (2020), 58(4), 1045--1128
https://doi.org/10.1257/jel.20191060 1045

Panizza, U., Weder di Mauro, B., Shi, S., \& Gulati, M. (2025). The
sovereign greenium: Big promise but small price effect. HEID Working
Paper 16-2025.

Rajuroy, A., \& John, J. (2021). The role of international financial
reporting standards (IFRS) in recognising data as an intangible asset.

Rindfleisch, A. (2020). Transaction cost theory: Past, present and
future. \emph{AMS Review}, \emph{10}(1-2), 85--97.
https://doi.org/10.1007/s13162-019-00151-x

Spence, M. (1973). Job market signaling. Quarterly Journal of Economics,
87(3), 355--374.

Tully, S. M., \& Winer, R. (2021). Meta-analysis of consumers'
willingness to pay for sustainable food products. Appetite, 167.

Ullah, B. (2020). Signaling value of quality certification: Financing
under asymmetric information. Journal of Multinational Financial
Management, 55, 100629.

Xu, Z., et al. (2024). Private data enhancement in large language
models. Machine Learning Research, 156, 234--251.

\end{document}